\documentclass[aps,prc,reprint,showpacs,superscriptaddress,twocolumn,notitlepage]{revtex4-1}

\usepackage{graphicx,color} 
\usepackage{bm}       
\usepackage{amsmath} 
\usepackage{amssymb}
\usepackage[dvipsnames]{xcolor}
\usepackage[normalem]{ulem}
\usepackage{siunitx}
\usepackage{multirow}
\usepackage{gensymb}
\usepackage{natbib}

\newcommand {\nc} {\newcommand}

\nc {\IR} [1]{\textcolor{red}{#1}}
\nc {\IB} [1]{\textcolor{blue}{#1}}
\nc {\IP} [1]{\textcolor{magenta}{#1}}
\nc {\IM} [1]{\textcolor{Bittersweet}{#1}}
\nc {\IE} [1]{\textcolor{Plum}{#1}}
\nc {\IC} [1]{\textcolor{cyan}{#1}}

\nc{\ninej}[9]{\left\{\begin{array}{ccc} #1 & #2 & #3 \\ #4 & #5 & #6 \\ #7 & #8 & #9 \\ \end{array}\right\}}
\nc{\sixj}[6]{\left\{\begin{array}{ccc} #1 & #2 & #3 \\ #4 & #5 & #6 \\ \end{array}\right\}}
\nc{\threej}[6]{ \left( \begin{array}{ccc} #1 & #2 & #3 \\ #4 & #5 & #6 \\ \end{array} \right) }
\nc{\half}{\frac{1}{2}}
\nc{\numberthis}{\addtocounter{equation}{1}\tag{\theequation}}
\nc{\lla}{\left\langle}
\nc{\rra}{\right\rangle}
\nc{\lrme}{\left|\left|}
\nc{\rrme}{\right|\right|}

\newcommand{\hw}{\ensuremath{\hbar\Omega}}

\newcommand{\SU}[1]{\ensuremath{\mathrm{SU}( #1 )}}

\newcommand{\SpR}[1]{\ensuremath{\mathrm{Sp}( #1,\mathbb{R} )}}

\begin{document}

\title{\textit{Ab initio} leading order effective potential for elastic proton scattering\\ based on the symmetry-adapted no-core shell model}

\author{R.~B.~Baker}
\affiliation{Institute of Nuclear and Particle Physics, and Department of Physics and Astronomy, Ohio University, Athens, OH 45701, USA}

\author{Ch.~Elster}
\affiliation{Institute of Nuclear and Particle Physics, and Department of Physics and Astronomy, Ohio University, Athens, OH 45701, USA}
\author{T.~Dytrych}
 \affiliation{Nuclear Physics Institute of the Czech Academy of Sciences, 250 68 \v{R}e\v{z}, Czech Republic}
\author{K.~D.~Launey}
\affiliation{Department of Physics and Astronomy, Louisiana State University, Baton Rouge, LA 70803, USA}

\date{\today}

\begin{abstract}
\begin{description}
\item[Background] 
Calculating microscopic optical potentials for elastic scattering at intermediate energies from light nuclei in an {\it ab initio} fashion within the Watson expansion has been established within the last few years. 
\item[Purpose] 
Based on the Watson expansion of the multiple scattering series, we
employ a nonlocal translationally invariant nuclear density derived
within the symmetry-adapted no-core shell model (SA-NCSM) framework from a
chiral next-to-next-to-leading order (NNLO) nucleon-nucleon interaction
and the very same interaction for a consistent
full-folding calculation of the effective (optical) potential for nucleon-nucleus scattering for medium-heavy nuclei.
\item[Methods] 
The leading order effective (optical) folding potential is computed by integrating over a translationally invariant SA-NCSM one-body scalar density, spin-projected momentum distribution, and the Wolfenstein amplitudes $A$, $C$, and $M$. The resulting nonlocal potentials serve as input for a momentum space Lippmann-Schwinger equation. 
In the SA-NCSM, the model space is systematically up-selected using $\SpR{3}$ symmetry considerations.

\item[Results] 
For the light nucleus of $^6$He, we establish a systematic selection scheme in the SA-NCSM for scattering 
observables. 
Then, we apply this scheme to calculations of scattering observables, such as differential cross sections, analyzing powers, and spin rotation functions for elastic proton scattering from $^{20}$Ne and $^{40}$Ca in the energy regime between 65 and 200~MeV, and compare to available data.

\item[Conclusions]
Our calculations show that the leading order effective nucleon-nucleus potential in the Watson expansion of multiple scattering theory obtained from an up-selected SA-NCSM model space describes $^{40}$Ca elastic scattering observables reasonably well to about 60 degrees in the center-of-mass frame, which coincides roughly with the validity of the NNLO chiral interaction used to calculate both the nucleon-nucleon amplitudes and the one-body scalar and spin nuclear densities.

\end{description}
\end{abstract}

\maketitle

\section{Introduction}
\label{sec:intro}

The study of atomic nuclei depends on nuclear reactions to extract  
structure and dynamics
observables. A specific approach to studying nuclear reactions consists of reducing the many-body scattering
problems to a few-body problem by isolating the relevant degrees of freedom~\cite{Johnson:2019sps} to arrive at
a few-body problem that is solved with the use of effective interactions, which are often called optical
potentials. While different techniques have been implemented for these effective interactions from first principles,
e.g. Refs.~\cite{Rotureau:2016jpf,Idini:2018qcj,Burrows:2023ygq,Burrows:2020qvu,Gennari:2017yez,Arellano:2022tsi},
we focus here on the use of the symmetry-adapted no-core shell model (SA-NCSM)~\cite{LauneyDD16,Dytrych:2020vkl,LauneyMD_ARNPS21}
to provide the relevant structure inputs. Specifically, we combine the one-body densities for the target
calculated within the SA-NCSM framework with the multiple scattering approach in leading order in the spectator
expansion to arrive at {\it ab initio} effective interactions for elastic nucleon-nucleus ($NA$) scattering. This
spectator expansion allows using the same nucleon-nucleon ($NN$) interaction when calculating the
one-body densities, which are then folded with those $NN$ amplitudes. Using realistic $NN$ (and
three-nucleon $3NF$ interactions) derived from chiral effective field theory, we can implement this procedure within a fully
\textit{ab initio} framework, provided we include all relevant terms in the spectator expansion at each order. The recent
work of Ref.~\cite{Burrows:2020qvu} has constructed and implemented effective nucleon-nucleus interactions that 
include the spin of the struck target nucleon consistently at the leading order.

The pioneering work in deriving an {\it ab initio} effective interaction for $NA$ elastic scattering for intermediate projectile energies was based
on the no-core shell model (NCSM) and thus limited to light nuclei with masses up to
$A\simeq16$~\cite{Burrows:2018ggt,Burrows:2020qvu,Baker:2023uzx,Gennari:2017yez} for reasonably well-converged
calculations of binding energies. The SA-NCSM can push the structure calculations to higher mass nuclei ($A \simeq$
48~\cite{LauneyMD_ARNPS21,Burrows:2023ugy}) by considering 
shape-related symmetries to construct the basis and selecting only the nonnegligible configurations.
The advantage of this selection process is
the drastic reduction in the number of basis states, which in turn allows calculations to move toward heavier nuclei. 

In this work, the non-local, translationally invariant scalar one-body densities and  spin-projected momentum
distributions  are derived from the SA-NCSM and employed for the calculation of a leading order -- in the spectator
expansion --  effective 
$NA$ interaction for targets of the halo nucleus $^6$He, the deformed $^{20}$Ne, and the medium-mass $^{40}$Ca nucleus. For the underlying $NN$ interaction used in the structure as well
as reaction calculation, we choose the 
chiral $NN$ interaction at the next-to-next-to-leading order
NNLO$_{\rm opt}$ from Ref.~\cite{Ekstrom13}.
This interaction is fitted with $\chi^2 \approx 1$ per degree of freedom for laboratory energies up to about
125~MeV. In the $A=3, 4$ nucleon systems, the contribution of three-nucleon forces ($3NF$s) of this interaction is
smaller than in most other parameterizations of chiral interactions. Consequently, nuclear quantities like
root-mean-square radii and electromagnetic transitions in light and intermediate-mass nuclei can be calculated
reasonably well without invoking $3NF$s~\cite{LauneyMD_ARNPS21,Henderson:2017dqc,Ruotsalainen19,LauneyMSSBMDD18,PhysRevC.106.024001}. In addition, observables calculated with the NNLO$_{\rm opt}$ $NN$ interaction have been found to be in good agreement with those calculated with other chiral potentials that require the use of the corresponding three-nucleon forces  (see, e.g., Refs. \cite{Burrows:2018ggt,BakerLBND20,Sargsyan:2021jme}).
From this point of view, the NNLO$_{\rm opt}$ chiral $NN$
interaction is very well suited for elastic scattering calculations using an optical potential based on the
 leading order in the spectator expansion since this order only contains explicit two-nucleon forces.
Other choices for structure methods and realistic nuclear interactions can be made, e.g., leading order optical potential calculations have also been performed for scattering from $^{40}$Ca in Ref.~\cite{Vorabbi:2023mml}
based on densities obtained from self-consistent Green's function using the NNLO$_{\rm sat}$ chiral interaction.

The structure of the paper is as follows. In Sec.~\ref{sec:theory}, we first review the basic approach for the
SA-NCSM, then we illustrate the selection prescriptions for the model spaces and review their effect on
structure observables. We also briefly review the derivation of the leading order effective $NA$ interaction
in calculating scattering.
In Sec.~\ref{sec:results}, we first show the effect of different symmetry-adapted selections for structure and proton elastic scattering observables from $^6$He as a test case.
The choice of this test case is motivated by the fact that $^6$He is a $p$-shell nucleus, for which highly
converged traditional NCSM calculations exist. Then, we apply those findings to elastic proton scattering from
$^{20}$Ne and $^{40}$Ca, and conclude in Sec.~\ref{sec:conclusions}.


\section{Theoretical Frameworks}
\label{sec:theory}

\subsection{Symmetry-Adapted No-Core Shell Model}
\label{sancsm}

The SA-NCSM is an \textit{ab initio} many-body approach that can achieve drastically reduced model spaces based on symmetries inherent to nuclei~\cite{DytrychLDRWRBB20}. This allows one to describe heavier nuclear systems and spatially expanded nuclear modes, including collective, clustering, and continuum degrees of freedom. The SA-NCSM framework is reviewed in Refs. \cite{LauneyDD16, LauneyMD_ARNPS21}.
An important feature of the symmetry-adapted (SA) framework is that the model space is reorganized to an SA basis that respects the deformation-related \SU{3} symmetry or the shape-related \SpR{3} symmetry \cite{LauneyDD16}. While the approach utilizes symmetry groups to construct the basis and the many-body Hamiltonian matrix (e.g., see Refs.~\cite{AkiyamaD73,DraayerLPL89,LangrDLD18,1937-1632_2019_0_183}), calculations are not limited {\it a priori} by any symmetry. They employ a large set of basis states that can describe a significant symmetry breaking if the nuclear Hamiltonian demands it. In addition, when necessary, the SA-NCSM calculations can be performed in complete model spaces that are equivalent within a unitary transformation to the ones used in NCSM.
Key features, especially the selection of nonnegligible contributions within the model space, are described in Ref.~\cite{LauneyDSBD20}.

The many-nucleon basis states of the SA-NCSM are labeled according to \SU{3}$_{(\lambda\,\mu)}\times$\SU{2}$_S$ by the total intrinsic spin $S$ and $(\lambda\,\mu)$ quantum numbers, in addition to many other quantum numbers needed to provide a complete labeling, including the nucleon distribution across the harmonic oscillator (HO) major shells, total proton spin and total neutron spin.  
Specifically, $\lambda=N_z-N_x$ and $\mu=N_x-N_y$, where $N_x+N_y+N_z$ is the total HO quanta distributed in the $x$, $y$, and $z$ direction. The \SU{3} quantum numbers describe deformation (see Ref. \cite{Heller:2022hhy}), and for example, the case of $N_x=N_y=N_z$, or equally $(\lambda\,\mu)=(0\, 0)$, describes a spherical configuration, while $N_z$ larger than  $N_x=N_y$ ($\mu=0$) indicates prolate deformation. A closed-shell configuration has $(0\, 0)$, so spherical modes (or no deformation) are a part of the SA basis. However, most nuclei, from light to heavy, are deformed in the body-fixed frame ($N_z > N_x > N_y$), which appear spherical in the laboratory frame for $0^+$ ground states.

We emphasize that within the SA-NCSM selected model spaces, the spurious center-of-mass motion can be exactly factored out from the intrinsic dynamics~\cite{Verhaar60,Hecht71} (see, e.g., Ref. \cite{Burrows:2023ygq}). This plays an important role in scattering calculations since the necessary one-body densities computed in the SA-NCSM are exactly translationally invariant (without any center-of-mass spuriousity). 

\subsection{A Selection Procedure for the SA Calculations}
\label{subsec:theory-select}

In the SA-NCSM, all basis states are kept up to a given $N$, while for higher $N$ ($N\leq N_{\rm max}$), the model space is systematically selected using \SpR{3} considerations (as in NCSM, the model space is truncated at $N_{\rm max}$ defined as the maximum number of HO quanta allowed in a many-particle state above the minimum for a given nucleus). Hence, the SA model spaces are labeled as ``$\left<N\right>N_{\rm max}$". Configurations that are highly favored in the $N$ model space inform important configurations in the $N+2$ model space, which in turn inform the $N+4$ model space, etc., and those track with larger deformation along the $N_z$ axis.  
Notably, these $N+4$ configurations can be readily reached from the $N+2$ configurations in the $N_z$-$N_x$ plane by two excitations in the $z$ direction.  

In this paper, we adopt a selection prescription, detailed in Ref.~\cite{LauneyDSBD20}, that has been heretofore tested for structure observables only.  Here, we apply it for the first time to scattering observables. Namely, we introduce a selection cutoff $\varepsilon_{\rm max}$, given by the fraction of the model space used, that is,
\begin{equation}
   \varepsilon=\frac{\dim_{\rm SA}(N_{\rm max})}{\dim(N_{\rm max})}\leq \varepsilon_{\rm max} \leq 1,
\label{eqn:eps}   
\end{equation}
where $\dim(N_{\rm max})$ is the dimensionality of the complete model space for a given $N_{\rm max}$ (and ``SA" denotes its selected counterpart).
The order in which basis states are included in the SA model space is determined according to the weight (see Ref.~\cite{LauneyDSBD20}),
\begin{eqnarray}
w(N_x,N_y,N_z+2)=\frac{P(N_x,N_y,N_z)}{\dim(N_x,N_y,N_z+2)}, 
\label{eqn:SA-weight}
\end{eqnarray}
where $P(N_x,N_y,N_z)$ is the probability amplitude of the eigenfunction obtained in SA-NCSM calculations in the smaller $N$ model space (e.g., the ground state, if this is the state of interest), and $\dim(N_x,N_y,N_z+2)$ denotes the dimensionality of the configuration in the larger $N+2$ model space to be selected (spin degrees are omitted for simplicity). The prescription is then applied to $N+4$ up through $N_{\rm max}$. 
For ease of comparing across different $N$ values (since configurations for large $N$ have much smaller probability amplitudes compared to those for low $N$), we normalize $w$ of Eq.~\eqref{eqn:SA-weight} to the highest weight value $w_{\max}$ in a given $N$:
\begin{eqnarray}
    w_{\mathrm{norm}}(N_x,N_y,N_z+2)=\frac{w(N_x,N_y,N_z+2)}{w_{\max}(N_x,N_y,N_z+2)}.
\end{eqnarray}

Similar to the NCSM, a measure of convergence for the results is the degree to which the SA-NCSM obtains results independent of the model parameters \hw~(the HO frequency), $N_{\rm max}$,  and $\varepsilon_{\rm max}$. Remarkably, even for small $\varepsilon_{\rm max}$ cutoffs, which correspond to drastically reduced model spaces, observables such as, e.g., B(E2) values are quite close to the converged results, a feature that further improves with $N_{\rm max}$ \cite{LauneyDSBD20}. 
In this paper, we show that the same 
selection scheme is valid for elastic scattering observables.

\subsection{The Leading Order Effective NA Interaction }

Calculating elastic nucleon-nucleus scattering observables in an {\it ab initio} fashion
requires not only the interaction between the nucleons within the target but also the interaction between the projectile and the nucleons in the target.
A multiple scattering expansion provides a framework to organize these interactions in a tractable way.
For example, the spectator expansion~\cite{Siciliano:1977zz,Baker:2023uzx} organizes the scattering of a nucleon from a nucleus consisting of $A$ nucleons
in terms of active nucleons. In the leading order of the spectator expansion, there are two active nucleons,
the projectile and one target nucleon. The next-to-leading order will have three active nucleons, the
projectile and two target nucleons, and so on. Thus, by construction, the leading order term only contains
the two-nucleon force between the projectile and the struck target nucleon. A scalar one-body density and a spin-projected momentum distribution represent the struck nucleon in the target, 
here calculated by employing {\it ab initio} many-body methods. For the current work, we use 
the SA-NCSM, which has been  applied
up to medium-mass nuclei, i.e., masses up to $A\approx 48$~\cite{LauneyMD_ARNPS21,
Burrows:2023ugy}. This nonlocal, translationally invariant one-body density~\cite{Burrows:2017wqn} is then folded with
off-shell $NN$ amplitudes given in the Wolfenstein
parameterization~\cite{wolfenstein-ashkin,Wolfenstein:1956xg}.
To ensure that the two-nucleon interactions are treated consistently in the structure and reaction
calculation, the spin of the struck target nucleon must be considered. This leads to a folding with the
well-known scalar one-body density matrix and a spin-projected one-body momentum
distribution. This ensures that central, spin-orbit, and tensor parts of the $NN$ interaction enter the effective $NA$ interaction.
We refer interested readers to Ref.~\cite{Burrows:2020qvu} for the formal derivation of the leading order $NA$ effective interaction.

For the densities considered in this work, we concentrate on proton scattering
from nuclei with  $J^\pi= 0^+$ in leading order in the spectator expansion.
In this case, the effective interaction of the proton projectile with a single target nucleon can be written as
a function of the momentum transfer ${\bf q}$ and the average momentum ${\bm{\mathcal{K}}_{NA}}$, where the
subscript $NA$ refers to the nucleon-nucleus ($NA$) frame. The effective $NA$ interaction in the leading order of the
spectator expansion is given as
\begin{widetext}
\begin{eqnarray}
\label{eq:uopt}
\lefteqn{\widehat{U}_{\mathrm{p}}(\bm{q},\bm{\mathcal{K}}_{NA};\epsilon) =} & &  \cr
& & \sum_{\alpha=\mathrm{n,p}} \int d^3{\mathcal{K}} \eta\left( \bm{q}, \bm{\mathcal{K}}, \bm{\mathcal{K}}_{NA} \right)
A_{\mathrm{p}\alpha}\left( \bm{q}, \frac{1}{2}\left( \frac{A+1}{A}\bm{\mathcal{K}}_{NA} - \bm{\mathcal{K}} \right); \epsilon
\right) \rho_\alpha^{K_s=0} \left(\bm{\mathcal{P}'}, \bm{\mathcal{P}}  \right) \cr
&+&i (\bm{\sigma^{(0)}}\cdot\hat{\bm{n}}) \sum_{\alpha=\mathrm{n,p}} \int d^3{\mathcal{K}} \eta\left( \bm{q},
\bm{\mathcal{K}}, \bm{\mathcal{K}}_{NA} \right)
C_{\mathrm{p}\alpha}\left( \bm{q}, \frac{1}{2}\left( \frac{A+1}{A}\bm{\mathcal{K}}_{NA} - \bm{\mathcal{K}} \right);
\epsilon
\right) \rho_\alpha^{K_s=0} \left(\bm{\mathcal{P}'}, \bm{\mathcal{P}}  \right) \cr
&+&i \sum_{\alpha=\mathrm{n,p}} \int d^3{\mathcal{K}} \eta\left( \bm{q}, \bm{\mathcal{K}}, \bm{\mathcal{K}}_{NA}
\right) C_{\mathrm{p}\alpha} \left( \bm{q}, \frac{1}{2}\left( \frac{A+1}{A}\bm{\mathcal{K}}_{NA} - \bm{\mathcal{K}}
\right); \epsilon \right) S_{n,\alpha} \left(\bm{\mathcal{P}'}, \bm{\mathcal{P}} \right) \cos \beta\cr
&+&i (\bm{\sigma^{(0)}}\cdot\hat{\bm{n}}) \sum_{\alpha=\mathrm{n,p}} \int d^3{\mathcal{K}} \eta\left( \bm{q},
\bm{\mathcal{K}}, \bm{\mathcal{K}}_{NA} \right)  (-i)
M_{\mathrm{p}\alpha} \left( \bm{q}, \frac{1}{2}\left( \frac{A+1}{A}\bm{\mathcal{K}}_{NA} - \bm{\mathcal{K}}
       \right); \epsilon \right) S_{n,\alpha} \left(\bm{\mathcal{P}'}, \bm{\mathcal{P}}  \right) \cos \beta,
\end{eqnarray}
\end{widetext}
where the subscript p indicates the projectile as being a proton. The energy $\epsilon$ is taken in the impulse
approximation as half of the projectile energy.
The momentum vectors in the problem are given as 
\begin{eqnarray}
\label{eq:2}
\bm{q} &=& \bm{p'} - \bm{p} = \bm{k'} - \bm{k}, \cr
\bm{\mathcal{K}_{NA}} &=& \frac{1}{2} \left(\bm{k'} + \bm{k}\right), ~~~~~{\bf{\hat{p}}}= \frac{1}{2} \left(\bm{p'} + \bm{p}\right) \cr
\hat{\bm{n}}&=&\frac{\bm{\mathcal{K}_{NA}} \times \bm{q}}{\left| \bm{\mathcal{K}_{NA}} \times
\bm{q}\right|} \cr
\bm{\mathcal{K}}& =& {\bf{\hat{p}}} +\bm{\mathcal{K}_{NA}}/{A} \cr
\bm{\mathcal{P}}&=& \bm{\mathcal{K}}+\frac{A-1}{A}\frac{\bm{q}}{2},  \cr
\bm{\mathcal{P'}}&=& \bm{\mathcal{K}}-\frac{A-1}{A}\frac{\bm{q}}{2}  .
\end{eqnarray}
The momentum of the incoming proton is given by $\bm{k}$, its outgoing momentum by $\bm{k'}$, the momentum
transfer by $\bm{q}$, and the average momentum $\bm{\mathcal{K}}_{NA}$. The struck nucleon in the target has an initial momentum $\bm{p}$ and a final momentum $\bm{p'}$ with average momentum ${\bf{\hat{p}}}$. When defining the integration variable $\bm{\mathcal{K}}$, the recoil of the nucleus is taking into account.
   The two quantities representing the structure of the nucleus are the scalar one-body density
$\rho_\alpha^{K_s=0} \left(\bm{\mathcal{P}'}, \bm{\mathcal{P}}  \right)$ and the
spin-projected momentum distribution $S_{n,\alpha} \left(\bm{\mathcal{P}'}, \bm{\mathcal{P}} \right)$.
Both distributions are nonlocal and translationally invariant.
Lastly, the term $\cos \beta$ in Eq.~(\ref{eq:uopt}) comes from projecting $\bm{\hat{n}}$ from the $NN$ frame
to the $NA$ frame. For further details, see Ref.~\cite{Burrows:2020qvu}. The term $\eta\left( \bm{q}, \bm{\mathcal{K}}, \bm{\mathcal{K}}_{NA} \right)$  is the M{\o}ller
factor~\cite{CMoller} describing the transformation from the $NN$ frame to the $NA$ frame.

The functions $A_{\mathrm{p}\alpha}$, $C_{\mathrm{p}\alpha}$, and $M_{\mathrm{p}\alpha}$ represent the $NN$ interaction through Wolfenstein
amplitudes. Since the incoming proton can interact
with either a proton or a neutron in the nucleus, the index $\alpha$ indicates the
neutron ($\mathrm{n}$) and proton ($\mathrm{p}$) contributions, which are calculated separately and then summed up.
Concerning the nucleus, the operator $i (\bm{\sigma^{(0)}}\cdot \hat{\bm{n}})$ represents the spin-orbit operator in the momentum space of
the projectile. As such, Eq.~(\ref{eq:uopt}) exhibits the
expected form of an interaction between a spin-$\frac{1}{2}$ projectile and a target nucleus in a $J=0$ state \cite{RodbergThaler}.

When calculating $NA$ elastic scattering amplitudes,
the leading order term of Eq.~(\ref{eq:uopt}) does not directly enter a Lippmann-Schwinger type integral
equation for the transition amplitude.
To obtain the Watson optical potential $U_p(\bm{q},\bm{\mathcal{K}}_{NA}; \epsilon)$,
an additional integral equation needs to be solved~\cite{Baker:2023uzx},
\begin{equation}
\label{eq:watson}
U_{\rm p} = \widehat{U}_{\rm p} - \widehat{U}_{\rm p} G_0 (E) P U_{\rm p},
\end{equation}
where, for simplicity, the momentum variables are omitted. Here, $G_0(E)$ is the free $NA$ propagator and $P$ a projector on the ground state. As pointed out in Ref.~\cite{Baker:2023uzx}, for solving the scattering problem the reference energy separating bound and continuum states is chosen such that the ground state energy is set to zero, implying that the energies referring to the target Hamiltonian in $G_0 (E)$ are excitation energies of the target.
For the proton-nucleus scattering calculations the Coulomb interaction between the projectile and the target is included using the exact momentum space formulation described in Ref.~\cite{Chinn:1991jb}. 


\section{Results and Discussion}
\label{sec:results}

\subsection{Examining the selection procedure with $^6$He observables}

To study the effect the symmetry-adapted selection procedure described in Section \ref{subsec:theory-select} has on reaction observables, we first examine a light nucleus, where calculations
in complete model spaces are currently available and can be used for validations.
As shown in the third column of Table \ref{tab:6He-selected}, the dimension of the $^6$He, $0^+_{\mathrm{gs}}$ model space grows by three orders of magnitude from $N_{\mathrm{max}}=4$ to $N_{\mathrm{max}}=12$ (from a dimension of less than $10^3$ to over $10^6$). This offers a good opportunity to explore the effect of different selection criteria in more detail. The dimensions of the model spaces resulting from various selection cutoffs $\varepsilon_{\mathrm{max}}$ values are shown in Table \ref{tab:6He-selected}, where we have adopted the notation $\left < N \right > N_{\mathrm{max}}$ to signify that the complete basis is included up to $N$ and SA selections are included from $N$ to $N_{\mathrm{max}}$, based on normalized weights $w_{\mathrm{norm}}$, as mentioned above.
This results in a selection process where we construct  $\left < N_{\mathrm{max}}-2 \right > N_{\mathrm{max}}$ SA model spaces consisting of
basis states in the $N_{\mathrm{max}}$ subspace
with 
$w_{\mathrm{norm}}$  weights from $1$ to $0.1$ ($w_{\mathrm{norm}}>10^{-1}$), $1$ to $0.01$ ($w_{\mathrm{norm}}>10^{-2}$), and so on.

\begin{table}[h]
\begin{center}
\begin{tabular}{l|c|r|c}
\multirow{2}{*}{$N_{\mathrm{max}}$}              &   $w_{\mathrm{norm}}$  &  \multirow{2}{*}{dimension} & \multirow{2}{*}{$\varepsilon_{\mathrm{max}}$}   \\
    & threshold  &   &   \\ \hline \hline
4   &   -   &   905 &   1 \\ \hline
\multirow{5}{*}{$\left<4\right>$6}  &     $10^{-1}$  &   1\,121  &   0.14   \\
    &  $10^{-2}$  &   2\,189  &   0.28 \\
    &  $10^{-3}$  &   3\,477  &   0.44 \\
    &  $10^{-4}$  &   5\,182  &   0.66 \\
    &  $10^{-5}$  &   5\,611  &   0.71 \\ \hline
6      &      -   &   7\,854   &   1      \\ \hline
\multirow{6}{*}{$\left<6\right>$8}  &     $10^{-1}$  &   8,083  &   0.16   \\
    &  $10^{-2}$  &   12\,972 &   0.26 \\
    &  $10^{-3}$  &   23\,064 &   0.47 \\
    &  $10^{-4}$  &   34\,624 &   0.70 \\
    &  $10^{-5}$  &   38\,969 &   0.79 \\
    &  $10^{-6}$  &   40\,131 &   0.81 \\ \hline
8      &      -   &   49\,248      &    1   \\ \hline
\multirow{7}{*}{$\left<8\right>$10}  &    $10^{-1}$  &   49\,742    &   0.20  \\
    &  $10^{-2}$  &   62\,898 &   0.26  \\
    &  $10^{-3}$  &   91\,417 &   0.37  \\
    &  $10^{-4}$  &   162\,196    &   0.66  \\
    &  $10^{-5}$  &   202\,435    &   0.83  \\
    &  $10^{-6}$  &   211\,522    &   0.86  \\
    &  $10^{-7}$  &   212\,039    &   0.87  \\ \hline
10  &  -     &   245\,082     &   1  \\ \hline
\multirow{6}{*}{$\left<10\right>$12}  &   $10^{-1}$  &   246\,043  &   0.24  \\
    &  $10^{-2}$  &   296\,552    &   0.29   \\
    &  $10^{-3}$  &   385\,576    &   0.38  \\
    &  $10^{-4}$  &   680\,062    &   0.66  \\
    &  $10^{-5}$  &   837\,611    &   0.82  \\
    &  $10^{-6}$  &   905\,918    &   0.88  \\ \hline
12  &   -   &   1\,024\,654   &   1   \\ \hline \hline
\end{tabular}
\end{center}
\caption{Model space dimensions for the $0^{+}$ ground state of $^6$He. The complete model space dimensions are provided and identified as $\varepsilon_{\mathrm{max}}=1$, consistent with Eq.~(\ref{eqn:eps}). Selected model space dimensions are also provided. This table uses the notation $\left<N\right>N_{\mathrm{max}}$, where $\left<N\right>$ signifies all contributions up to $N$
are included (here $N=N_{\mathrm{max}}-2$), and SA selections are made in model spaces from $N+2$
to $N_{\mathrm{max}}$. The $\varepsilon_{\mathrm{max}}$ values describe different model spaces based on a probability weight. See text for further discussion.
}
\label{tab:6He-selected}
\end{table}

As can be seen in Table \ref{tab:6He-selected}, each model space may have a different range of selectable $w_{\mathrm{norm}}$ values. For example, the $\left < 4\right> 6$ model space has included all $N+2$ configurations by $w_{\mathrm{norm}}=10^{-5}$, but the $\left < 6\right> 8$ model space has $N+2$ configurations with normalized weights as small as $10^{-6}$, though their contributions to the results shown later are negligible. Note that the difference in
basis dimension from, e.g., $w_{\mathrm{norm}}>10^{-5}$ in $\left <4\right>6$ to $\left <6\right>$ comes from new configurations at $N_{\mathrm{max}}=6$ that are not connected to those in
$N_{\mathrm{max}}=4$
through the prescription of Eq. \eqref{eqn:SA-weight}.

\begin{figure}
\begin{center}
\begin{tabular}{c}
\includegraphics[width=8.6cm]{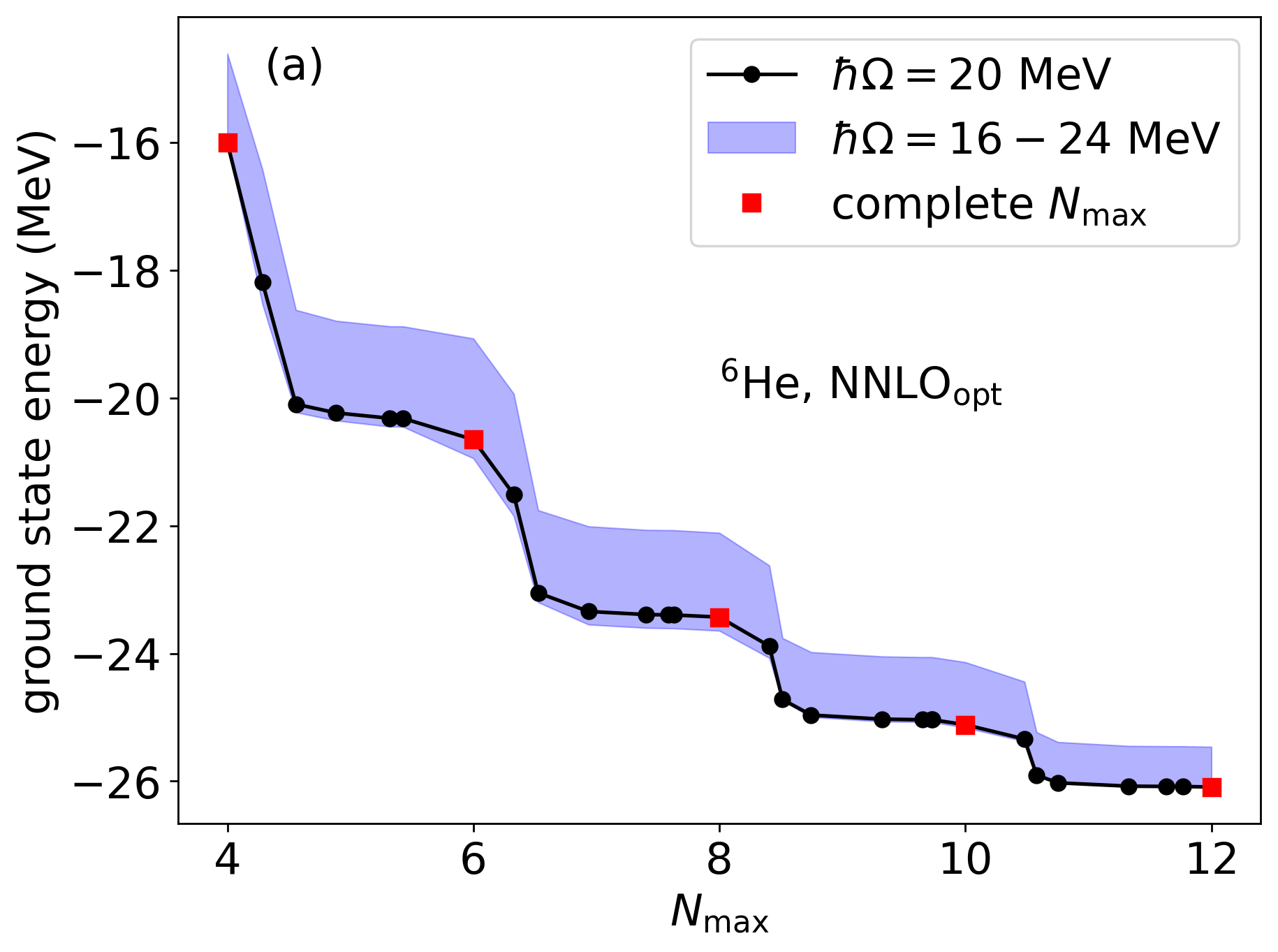} \\
\includegraphics[width=8.6cm]{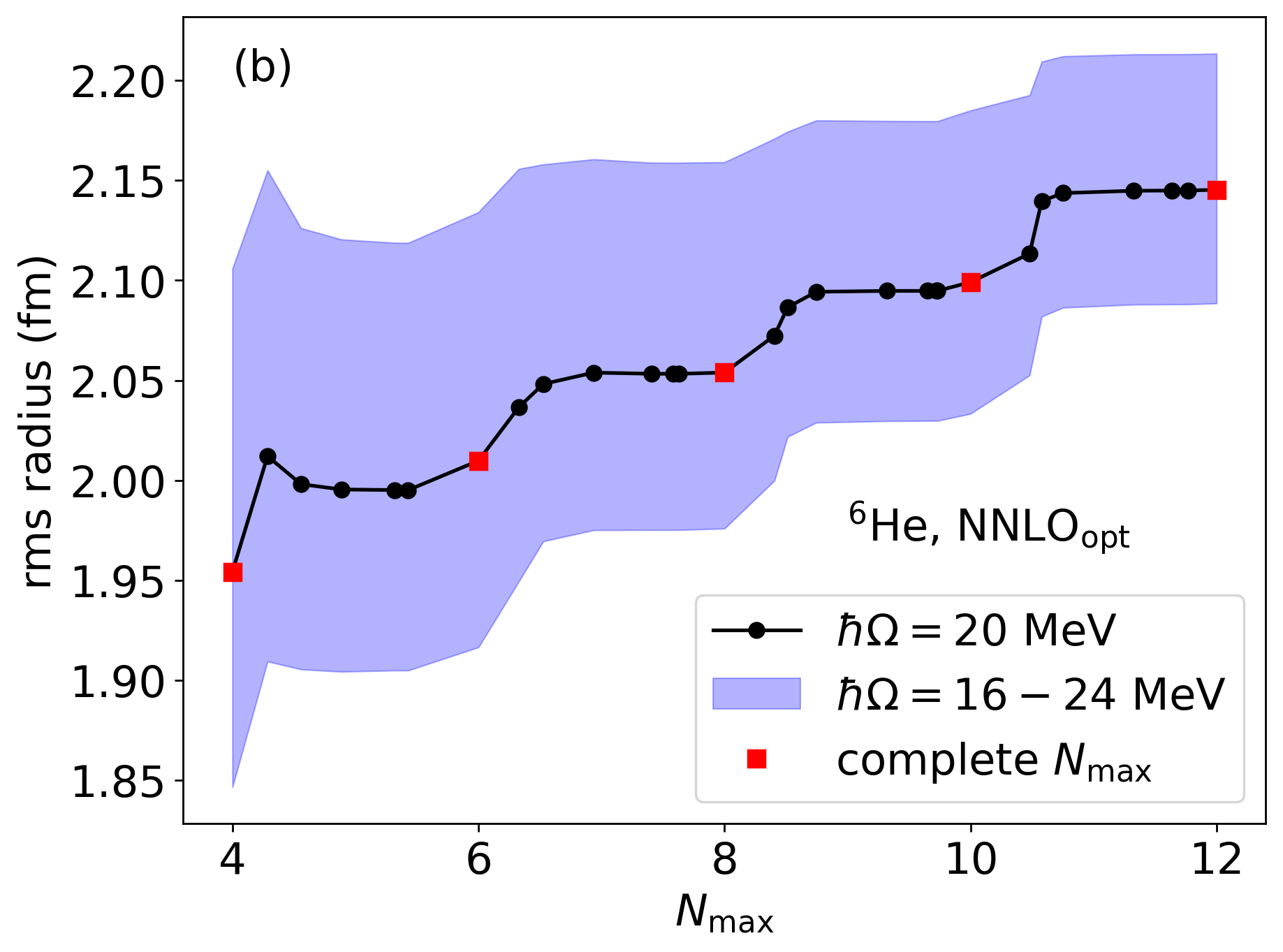}	\\
\includegraphics[width=8.6cm]{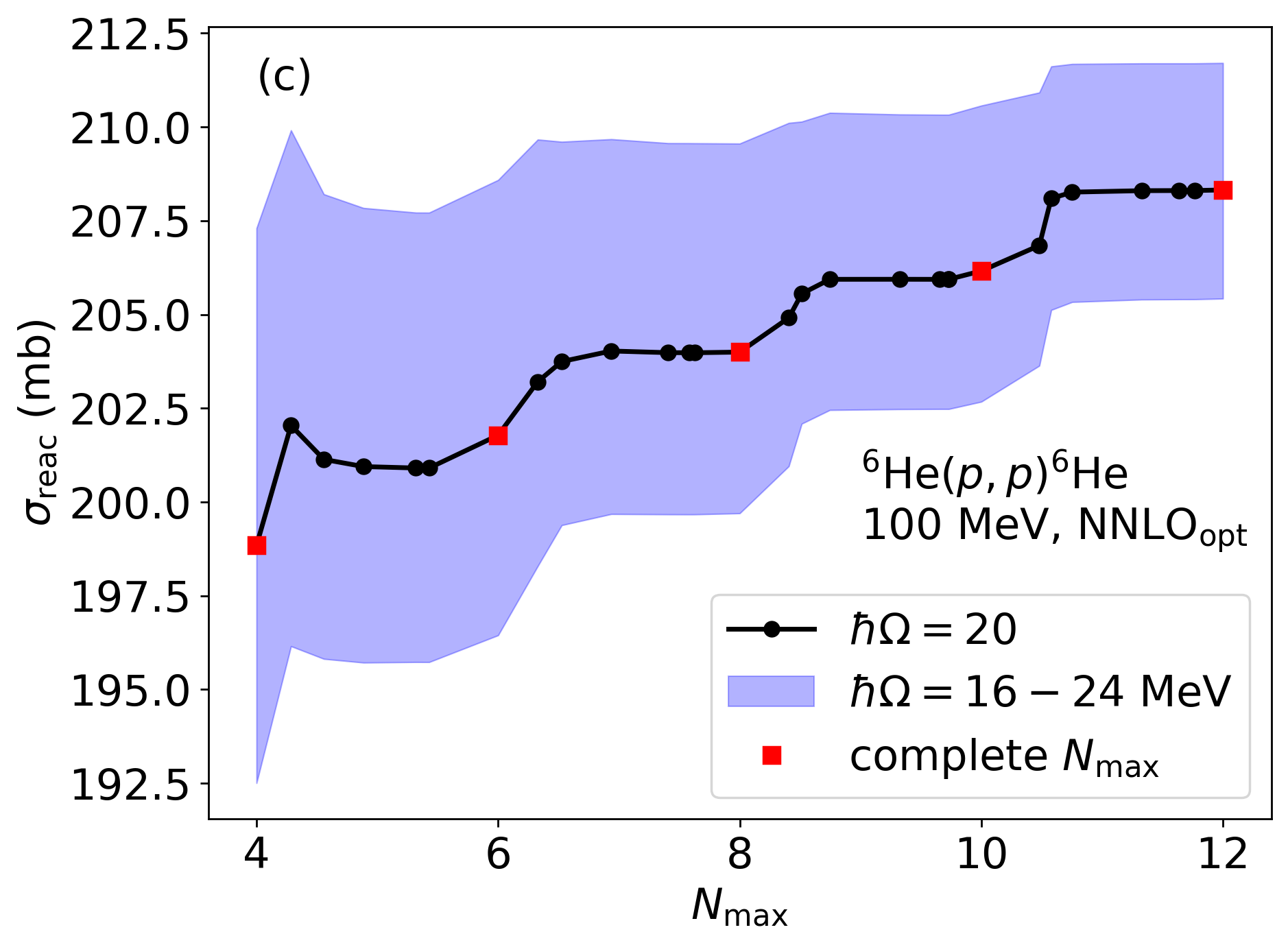}	\\
\end{tabular}
\caption{Plots of (a) the ground state energy of $^6$He, (b) the rms matter radius of the ground state of $^6$He, and (c) the reaction cross section for proton scattering on $^6$He at 100 MeV laboratory projectile kinetic energy. All results are shown as a function of $N_{\mathrm{max}}$, where the red squares correspond to calculations performed with complete model spaces, and the black dots correspond to calculations performed with SA-selected model spaces given by $\varepsilon_{\rm max}$, both at $\hbar\Omega=20$ MeV. The bands indicate differences between $\hbar\Omega=16-24$ MeV. See text for further discussion. 
}
\label{fig1}
\end{center}
\end{figure}

Using these SA model spaces in a structure calculation, the value of the corresponding structure observables is shown in Fig.~\ref{fig1}, with Fig.~\ref{fig1}(a) showing the ground state energy, and Fig.~\ref{fig1}(b) showing the root-mean-square (rms) matter radius.  
The red squares correspond to calculations in the complete model spaces, while the black dots correspond to symmetry-adapted model spaces at different $\varepsilon_{\mathrm{max}}$ values. The black dots are placed along the $x$-axis such that they indicate the percent of the model space included, e.g., a black dot near $N_{\mathrm{max}}=5$ corresponds to a $\left <4\right> 6$ model space with a dimension roughly $50\%$ the size of the complete $N_{\mathrm{max}}=6$ model space. Note that the bands indicate the variation in the results at nearby $\hbar\Omega$ values. The line for the center $\hbar\Omega$ value was selected according to $\hbar\Omega \approx 41/A^{\frac{1}{3}}$, 
which typically yields the fastest convergence of rms radii.
For $^6$He, this corresponds to $\hbar\Omega\approx20$ MeV, which also emerges as the variational minimum in the ground state energy as $N_{\mathrm{max}}$ increases [Fig.~\ref{fig1}(a)]. 

As shown by the structure observables in Fig.~\ref{fig1}(a) and Fig.~\ref{fig1}(b), in most cases, only a third of the model space ($\varepsilon_{\rm max}\sim 0.26$-$0.29$, constructed from $w_{\mathrm{norm}}>10^{-2}$) is already sufficient to reproduce the results of the complete model space.
This is particularly true at the larger $N_{\mathrm{max}}$ values. Notably, this pattern continues when examining scalar reaction observables. Namely, the reaction cross section for proton scattering at 100 MeV laboratory energy, $\sigma_{\mathrm{reac}}$, shown in Fig.~\ref{fig1}(c), 
has a convergence pattern almost identical to that of the rms radius. From each of these results, it is worth noting that the variation in each observable with respect to $\hbar\Omega$ is larger than the variation with respect to the model space selection -- that is, the width of the bands is larger than the difference between neighboring black points in Fig.~\ref{fig1}.

\begin{figure}
\begin{center}
\begin{tabular}{c}
\includegraphics[width=8.6cm]{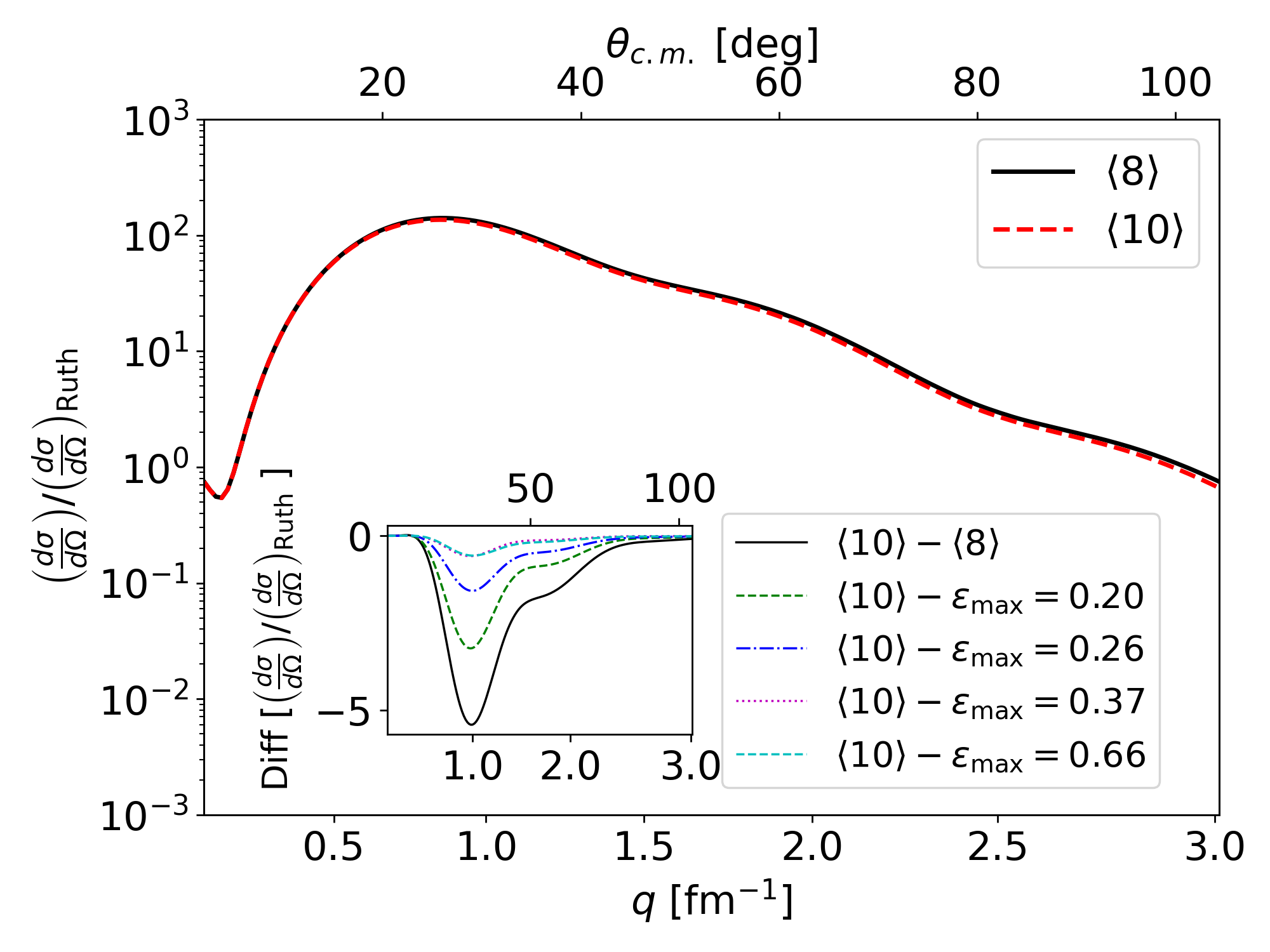}	\\ 
\includegraphics[width=8.6cm]{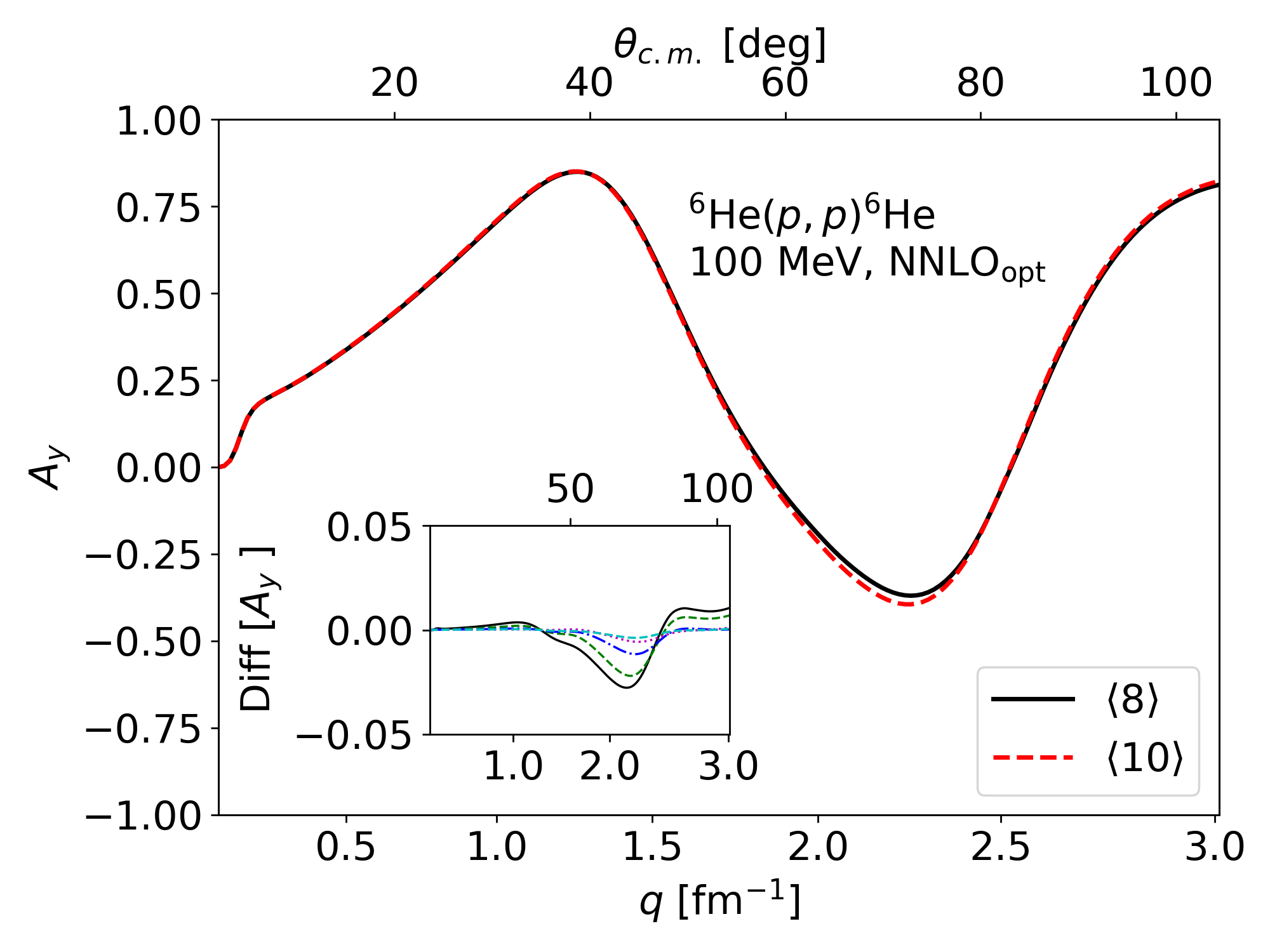} \\
\end{tabular}
\caption{Elastic scattering observables for $^6$He$(p,p)^6$He at 100 MeV laboratory energy. The top plot shows the differential cross section divided by the Rutherford cross-section, and the bottom plot shows the analyzing power $A_y$. Both plots illustrate the difference between $N_{\mathrm{max}}=8$ and $N_{\mathrm{max}}=10$ calculations -- labeled as $\left < 8 \right>$ and $\left < 10 \right>$, respectively -- with insets showing the differences between the different SA-selected model spaces given by $\varepsilon_{\rm max}$. The formatting is the same across both insets. See text for further discussion.}
\label{fig2}
\end{center}
\end{figure}

\begin{figure}
\begin{center}
\begin{tabular}{c}
\includegraphics[width=8.6cm]{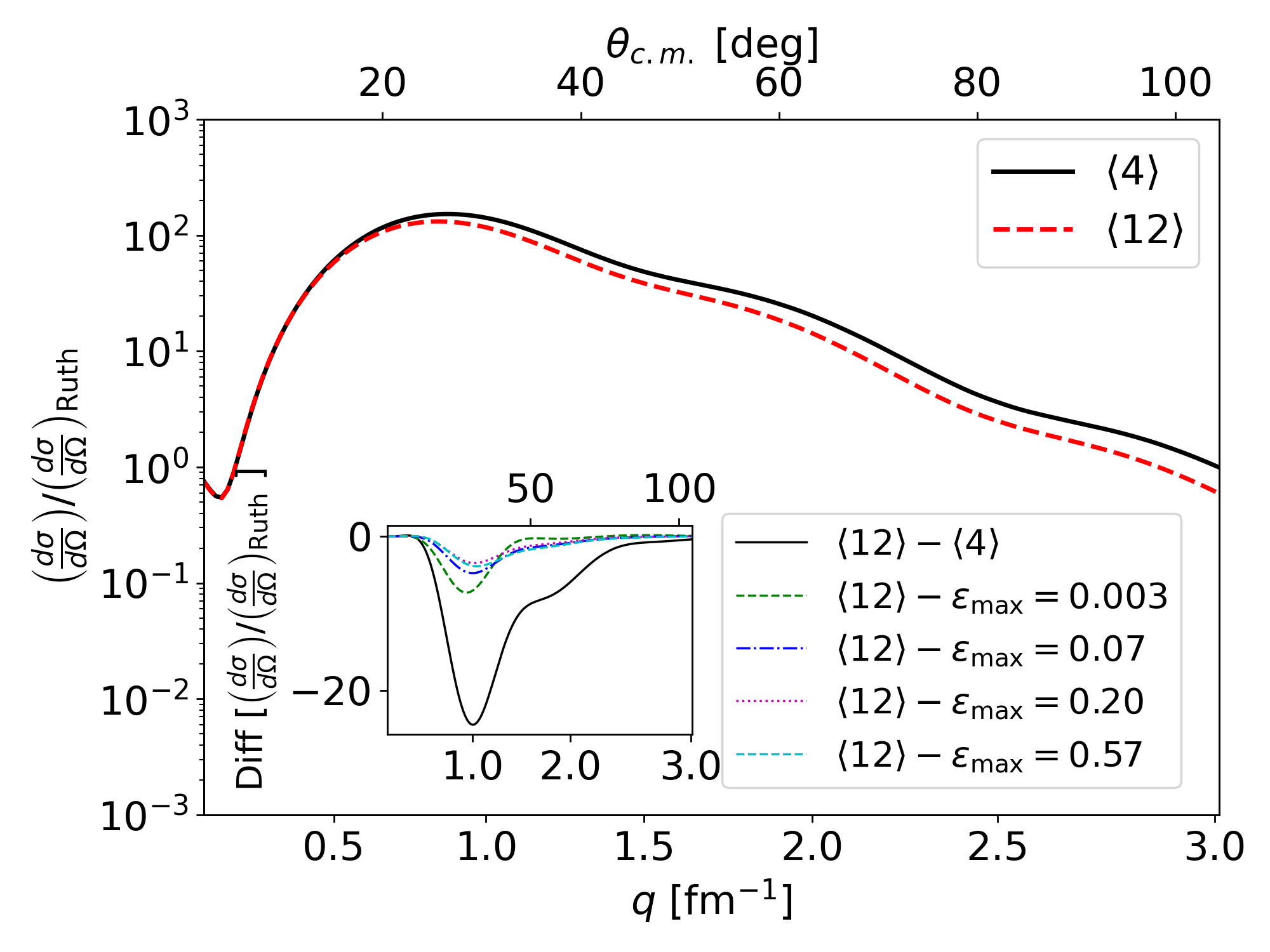}	\\ 
\includegraphics[width=8.6cm]{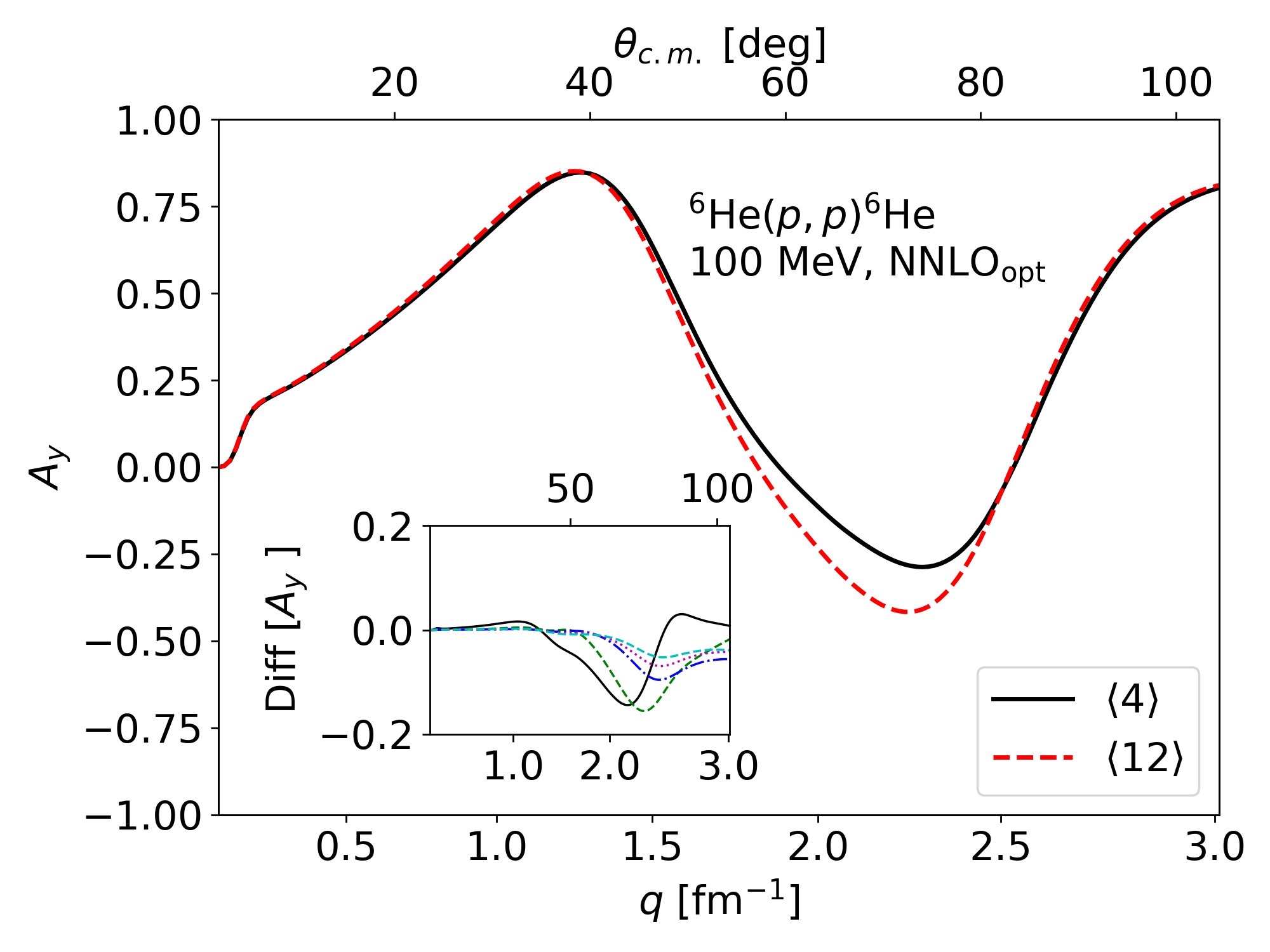} \\
\end{tabular}
\caption{Elastic scattering observables for $^6$He$(p,p)^6$He at 100 MeV laboratory energy. The top plot shows the differential cross-section divided by the Rutherford cross-section, and the bottom plot shows the analyzing power $A_y$. Both plots illustrate the difference between $N_{\mathrm{max}}=4$ and $N_{\mathrm{max}}=12$ calculations -- labeled as $\left < 4 \right>$ and $\left < 12 \right>$, respectively -- with insets showing the differences between the different SA-selected model spaces given by $\varepsilon_{\rm max}$. The formatting is the same across both insets. See text for further discussion.}
\label{fig2b}
\end{center}
\end{figure}

With an understanding of how the scalar observables converge for different model space selections, we can also examine functional observables, as shown in Figs.~\ref{fig2} and \ref{fig2b}, which shows the differential cross section and analyzing power for proton scattering on $^6$He at 100 MeV. Comparing the $N_{\mathrm{max}}=8$
and $N_{\mathrm{max}}=10$
results (Fig.~\ref{fig2}), small differences can be seen at large momentum transfers $q$, where previous work has already shown these observables are slower to converge with respect to $N_{\mathrm{max}}$~\cite{Burrows:2018ggt}. Focusing on these two $N_{\mathrm{max}}$ results, the inset shows the differences, where the solid black line shows how the results change from $N_{\mathrm{max}}=8$ to $N_{\mathrm{max}}=10$, and the other lines show the differences in the $N_{\mathrm{max}}=10$ and a selection of the $\left <8\right>10$ model spaces. Similar to the structure observables, by approximately  
$w_{\mathrm{norm}}>10^{-2}$ (or $\varepsilon=0.26$, roughly $26\%$ of the model space),
the differences for the scattering observables in the SA and complete spaces are quite small. While not shown here, the convergence pattern for the spin rotation function $Q$ is the same as the analyzing power $A_y$.

Similarly, Fig.~\ref{fig2b} shows the same observables but compares against a broader span of $N_{\mathrm{max}}$, namely $N_{\mathrm{max}}=4$ and $N_{\mathrm{max}}=12$. In this case, more significant differences are noticeable, as the $N_{\mathrm{max}}=12$ results are much closer to convergence than $N_{\mathrm{max}}=4$. We see similar behavior as in the previous examples by applying the same process to construct SA model spaces. Namely, a well-constructed SA model space can reproduce the complete space results to within a few percent, 
which is typically smaller than the
uncertainty from the $\hbar\Omega$ variation
or choice of the realistic interaction~\cite{Burrows:2018ggt, Baker:2023wla}. 

\subsection{Applying the selection procedure to proton scattering from $^{20}$Ne and $^{40}$Ca}

\begin{figure}
\begin{center}
\includegraphics[width=8.6cm]{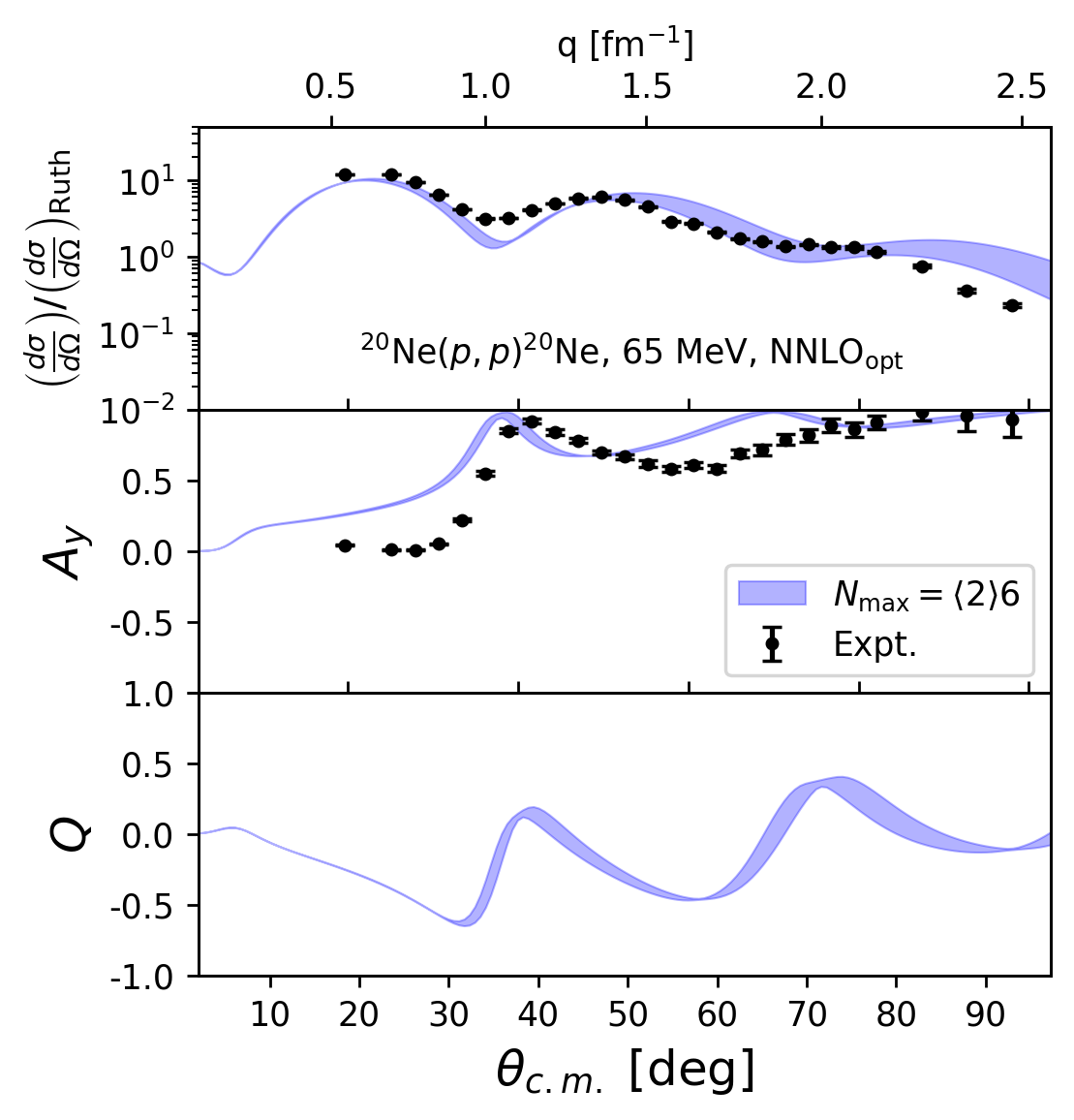}	
\caption{
The angular distribution of the differential cross section $\left(\frac{d\sigma}{d\Omega}\right)$ divided by the Rutherford cross-section, the analyzing power ($A_y$), and spin rotation function ($Q$) for elastic
proton scattering from $^{20}$Ne at 65~MeV laboratory projectile energy. 
The calculations use the NNLO$_{\rm opt}$ interaction in the SA-NCSM with $\langle 2 \rangle 6$ model spaces and $\varepsilon_{\rm max}=0.06$, and in 
the leading order in the effective $NA$ interaction. 
The band indicates
the differences between calculations performed with $\hbar\Omega$ from 13 to 17 MeV.
The data are taken from Ref.~\cite{Sakaguchi:1979fpk}.
}
\label{fig4}
\end{center}
\end{figure}

Acknowledging the convergence behavior of the observables for $^6$He as discussed in the previous section,
we now turn to heavier nuclei, namely, the doubly open-shell nucleus $^{20}$Ne and the closed-shell $^{40}$Ca, where 
complete model spaces for sufficiently large $N_{\rm max}$ are not feasible
due to the rapid growth of the model space dimensions. Applying the same selection procedure for
 the ground state of $^{20}$Ne, we construct a $\left < 2 \right >6$ model space and compute the
nonlocal scalar densities $\rho_\alpha^{K_s=0}$ and the spin-projected momentum distributions $S_{n,\alpha}$
(where $\alpha$ refers to the separate proton or neutron distributions) that enter the expression
$\widehat{U}_{\mathrm{p}}(\bm{q},\bm{\mathcal{K}}_{NA};\epsilon)$ of Eq.~(\ref{eq:uopt}) for the effective proton-nucleus interaction. Since experimental information for proton scattering from $^{20}$Ne in the energy regime above
$\sim$60~MeV is limited, we show in Fig.~\ref{fig4} elastic scattering observables at 65~MeV calculated with three different values of the oscillator parameter $\hbar\Omega$ indicated by the shaded band. The magnitude of the differential cross section (divided by the Rutherford cross section) is slightly underpredicted by the calculation, as is the first diffraction minimum. This small shift in the minimum corresponds to a slightly smaller rms matter
radius from the theory calculation -- here, $2.6(1)$~fm -- compared to the experimental value of $2.87(3)$~fm \cite{OZAWA200132}. This is consistent with the slightly smaller rms matter radii obtained by NNLO$_{\rm opt}$ in light nuclei~\cite{Burrows:2018ggt}. Additionally, the calculation deviates from the data for momentum transfers larger than 2~fm$^{-1}$. However, 
at the higher momentum transfers (or larger angles), the leading order in the spectator expansion should not be expected to be sufficient. Similarly, in addition to the slight radius discrepancy, rescattering terms may contribute to
the first minimum in the differential
cross section, as studies in few-body systems within the Faddeev framework~\cite{Elster:2008yt} suggest.
The calculation of the analyzing power $A_y$ follows the general shape of the data but does not describe them well. This should
also be taken as an indication that this energy is at the lower limit of applicability of the leading order term in the spectator expansion.
The spin rotation function is shown as a prediction, as no experimental data exists. 

\begin{figure}
\begin{center}
\includegraphics[width=8.6cm]{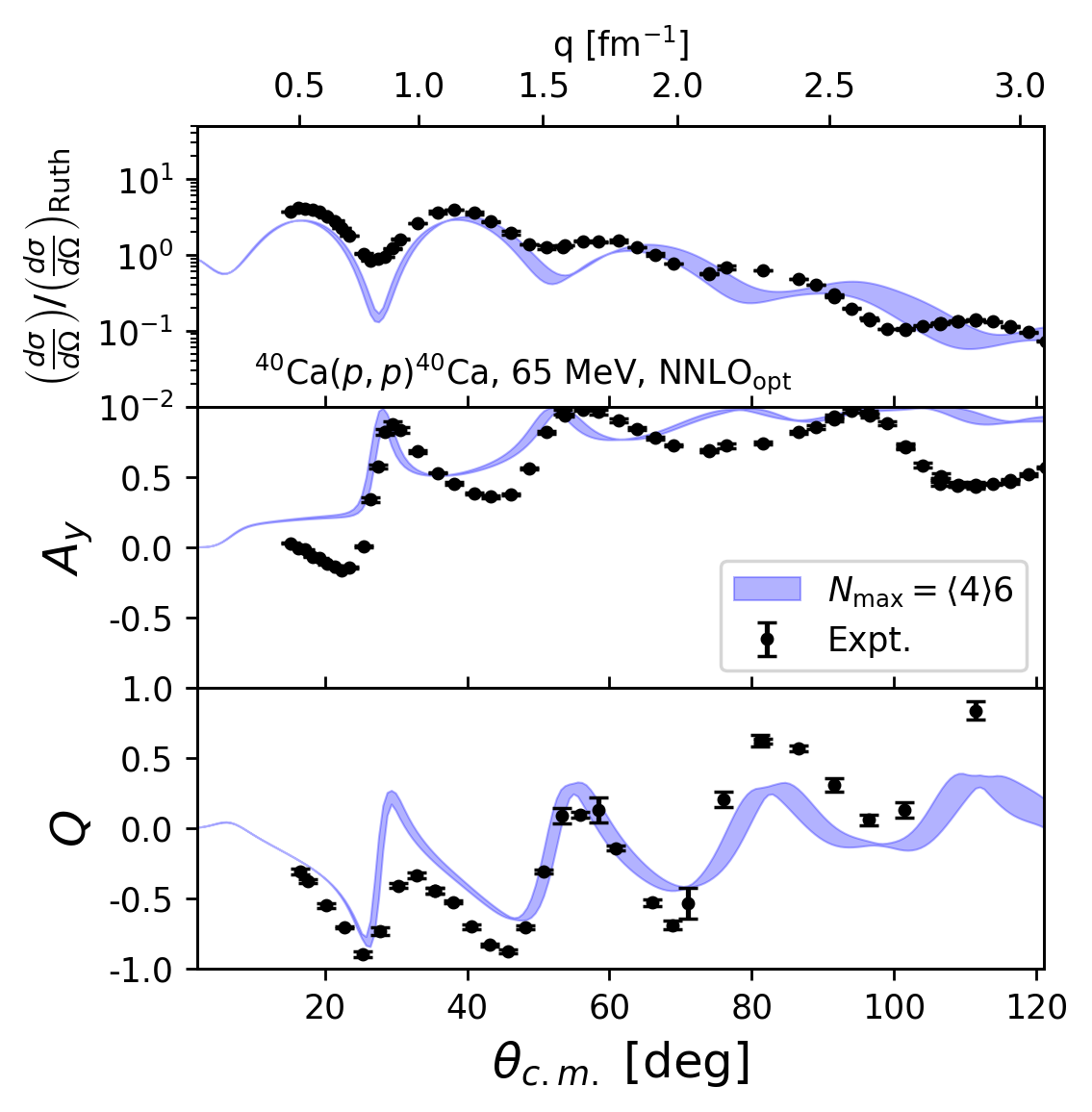}	
\caption{
The angular distribution of the differential cross section $\left(\frac{d\sigma} {d\Omega}\right)$
divided by the Rutherford cross section, analyzing power ($A_y$), and spin rotation function ($Q$) for elastic
proton scattering from $^{40}$Ca at 65~MeV laboratory projectile energy. The calculations use the NNLO$_{\mathrm{opt}}$ interaction in the SA-NCSM with $\left < 4\right > 6$ model spaces and $\varepsilon_{\mathrm{max}} = 0.03$ (constructed from $w_{\mathrm{norm}}>10^{-2}$), and in the leading order in the effective $NA$ interaction. The band indicates the difference between calculations performed with $\hbar\Omega$ between 11 and 13 MeV. The cross-section and $A_y$ data 
are taken from Ref.~\cite{Sakaguchi:1979fpk} and those for the spin rotation function from Ref.~\cite{Sakaguchi:1986}.
}
\label{fig5}
\end{center}
\end{figure}

Turning to $^{40}$Ca, we employ the same procedure for computing the nonlocal scalar densities and the spin-projected
momentum distributions. Unlike the dimensions of the $^6$He model spaces, the $^{40}$Ca model spaces grow significantly
faster, with the complete $N_{\mathrm{max}}=6$ space for $J=0$  having a dimension of 327\ 125\ 599. As a result, the important
states provided by the same $w_{\mathrm{norm}}$ values ($w_{\mathrm{norm}}>10^{-2}$) give smaller
$\varepsilon_{\mathrm{max}}$ values (a model space fraction of only $3\%$) than for $^6$He. Elastic proton scattering from
$^{40}$Ca is well measured in the energy regime between 65 and 200~projectile energy. Figure~\ref{fig5} shows the
calculations for $\hbar\Omega$ values between 11 and 13~MeV at 65~MeV laboratory projectile energy
  compared to the experimental data. In contrast to the
calculations for $^{20}$Ne, the underpredection of the differential cross section for small momentum transfers is larger, 
while the first diffraction minimum corresponds to the experimental one. Considering the diffraction pattern given by the first few minima, we see that it is wider than the experiment suggests. This may be related to the smaller calculated rms radius than the measured one. For the model spaces used here, this corresponds to an rms matter radius of $3.1(1)$~fm, compared to the experimental charge radius of $3.4776(19)$~fm \cite{ANGELI201369}.

Comparing 
the differential cross section for $^{40}$Ca, Fig.~\ref{fig5}, to the one for $^{20}$Ne at the same energy of 65 MeV, Fig.~\ref{fig4}, we observe that the description of the Neon data at higher momentum transfer is better. This may be related to $^{20}$Ne being a deformed, doubly open-shell nucleus, sometimes considered to have an $^{16}$O core with two extra protons and neutrons in the outer shell (see, e.g., Ref. \cite{DreyfussLESBDD20} for the projection of the $^{20}$Ne ground state on the s-wave $^{16}$O$+\alpha$, and Ref. \cite{LauneyMD_ARNPS21} for the cluster substructure revealed in the one-body density profile). If the nuclear density probed with proton scattering is less dense, rescattering, i.e., the next order in the multiple scattering expansion, may contribute less. Therefore, the leading order term gives a better description of the data. A similar effect has been seen in Ref.~\cite{Burrows:2020qvu} in the very good description of the differential cross section for proton scattering from $^6$He and $^8$He.

At 65~MeV, the analyzing power $A_y$ and the spin rotation function $Q$ are also measured for proton scattering from $^{40}$Ca. For small momentum transfers, $A_y$ is overpredicted by the calculation, while $Q$ is more consistent with the data. As mentioned earlier, 65~MeV projectile energy is at the lower limit of the validity of the leading order in the spectator expansion, and corrections to the leading order should become visible. In Ref.~\cite{Chinn:1993zz}, a modification of the many-body propagator due to the nuclear medium was introduced in a mean-field framework. In this work, calculations of the same observables for $^{40}$Ca show that those modifications improve the description of the spin-observables at 65~MeV while having no effect at higher energies. 

\begin{figure}
\begin{center}
\includegraphics[width=8.6cm]{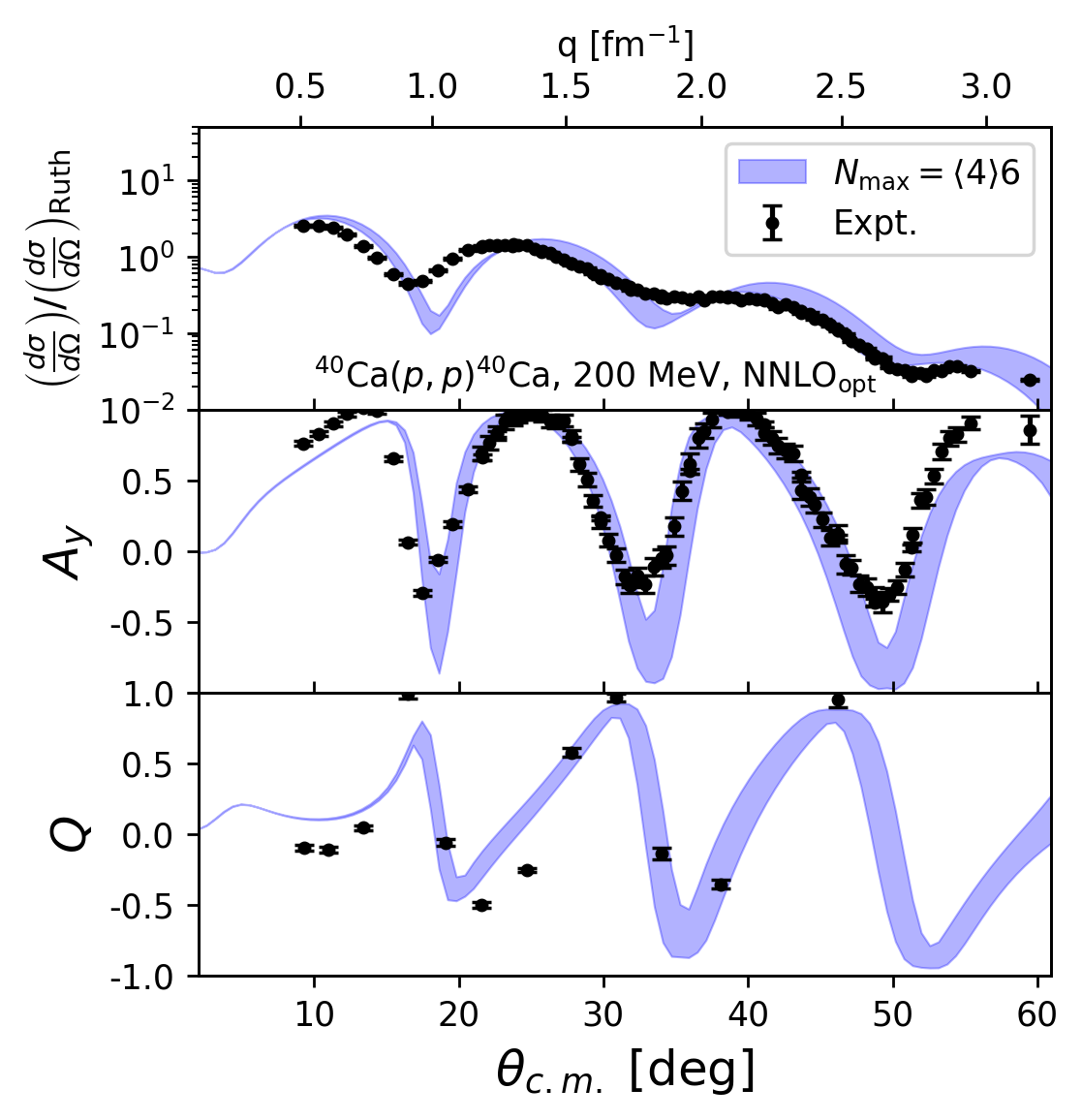}
\caption{
Same as Fig.~\ref{fig5} for proton scattering from $^{40}$Ca at 200~MeV laboratory projectile energy.  The cross-section and $A_y$ data
are taken from Refs.~\cite{Stephenson1985,Seifert:1993zz} and those for the spin rotation function 
from Ref.~\cite{Stephenson:1985Qca}.
}
\label{fig6}
\end{center}
\end{figure}

In Fig.~\ref{fig6}, the observables for elastic proton scattering from $^{40}$Ca for 200~MeV projectile energy are shown
for the same variation of $\hbar\Omega$ and the same model space. Here, the pattern of both spin observables is very well described. The differential cross section is slightly overpredicted for small momentum transfers, and the first calculated diffraction minimum is shifted toward higher momentum transfers. Only the next minima line up better with the experiment. This slight shift of the first minimum is also seen in $A_y$ and $Q$.

\begin{figure}
\begin{center}
\includegraphics[width=8.6cm]{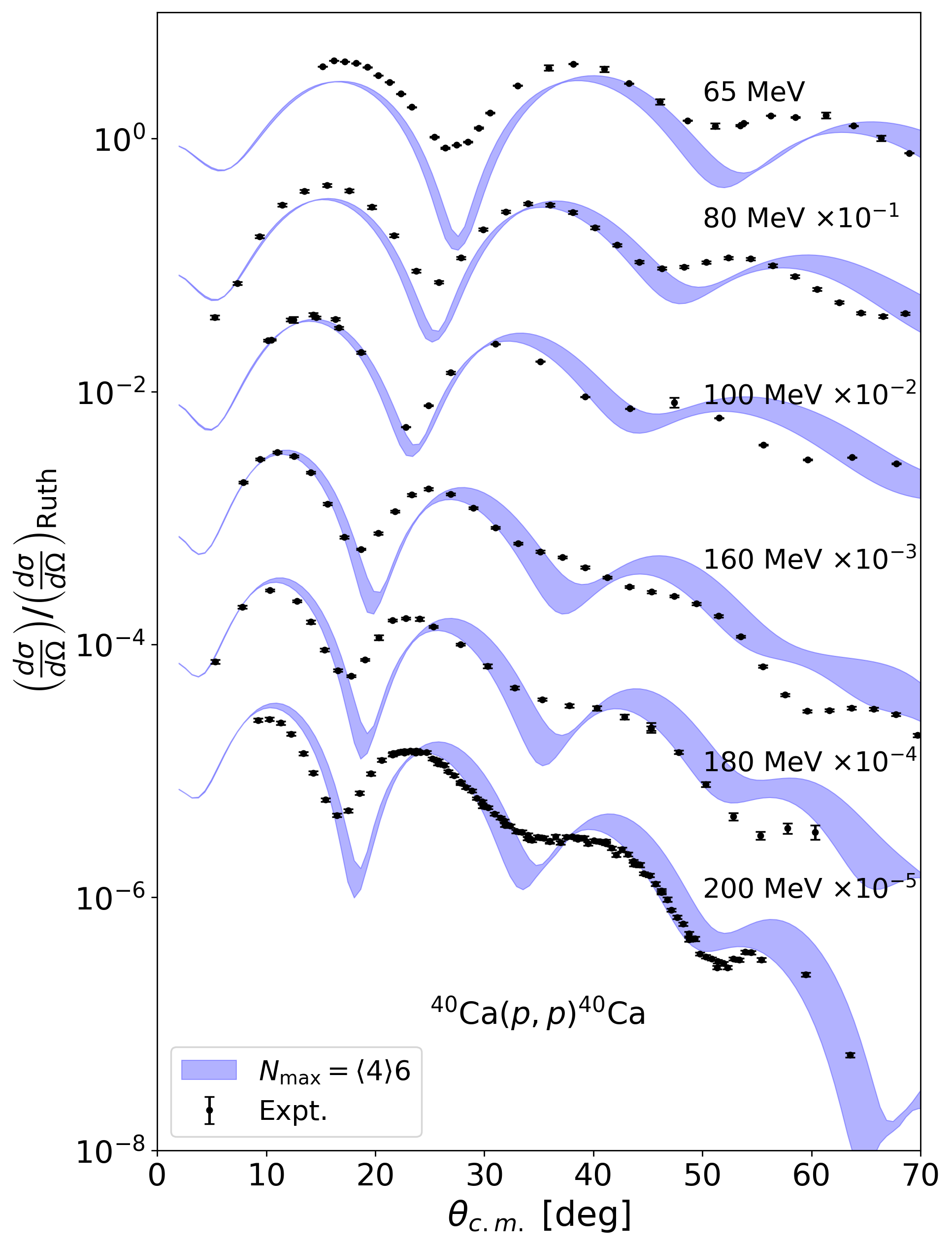}	
\caption{
The angular distribution of the differential cross section $\left(\frac{d\sigma} {d\Omega}\right)$
divided by the Rutherford cross section for elastic
proton scattering on $^{40}$Ca from 65 to 200~MeV laboratory projectile energy. The cross sections are multiplied
by the powers of 10 indicated at the energies listed in the figure.
The calculations use the NNLO$_{\rm opt}$ interaction in the SA-NCSM with $\langle 4 \rangle 6$ model spaces and $\varepsilon_{\rm max}=0.03$ (constructed from
$w_{\mathrm{norm}} > 10^{-2}$), and in
the leading order in the effective $NA$ interaction. 
The band indicates variations in the results from $\hbar\Omega=11-13$ MeV. The cross-section data
are taken from Ref.~\cite{Sakaguchi:1979fpk} for 65~MeV, Ref.~\cite{Nadasen:1981ep} for 80 and 160~MeV,
Ref.~\cite{Seifert:1990} for 100~MeV, Ref.~\cite{vanOers:1971es} for 180~MeV, and
Refs.~\cite{Stephenson1985,Seifert:1993zz} for 200~MeV.
}
\label{fig7}
\end{center}
\end{figure}

To study the energy dependence of elastic scattering observables, we show in Fig.~\ref{fig7} the differential cross section divided by the Rutherford cross section between 65 and 200~MeV projectile energy. The bands indicate the variation in $\hbar\Omega$ for the many-body structure calculations. The dependence on $\hbar\Omega$ observed for large angles suggests that larger model spaces may be needed to describe the data in this region. 
It is interesting to observe that 
for $100$ MeV, the overall agreement for $q \leq 1.5$ fm$^{-1}$ is remarkable, whereas the deviations at other energies may stem from rescattering effects not included at leading order (at lower energies) and properties of the $NN$ interaction (at higher energies). Indeed, 
for the energies lower than 100~MeV, the differential cross section for small momentum transfers is slightly underpredicted, while for energies higher than 180~MeV, the experiment is overpredicted. 
This could point to an issue with the energy dependence of leading order term of the spectator expansion derived
from the $NNLO_{\rm opt}$ interaction,
since for small momentum transfers rescattering effects should be small.
In addition, while for the lower energies, the location of the first minimum corresponds exactly to the experimentally observed one, at energies higher than 100~MeV, the calculated first minimum shifts towards larger angles. This shift of the first diffraction minimum towards higher angles as a function of projectile energy was also observed in Ref.~\cite{Vorabbi:2023mml}. Though in Born approximation and treating the nucleus as a black disk, the first minimum is directly related to the radius of the nucleus, the full calculation reveals an energy dependence of the location of the first minimum. Considering that the leading order term in the spectator expansion dominates the elastic scattering at higher energies, the predicted first minimum at 200~MeV being shifted to slightly larger angles is most likely related to the NNLO$_{\rm opt}$ properties that are responsible for underpredicting  
the rms radius.

\begin{figure}
\begin{center}
\includegraphics[width=8.6cm]{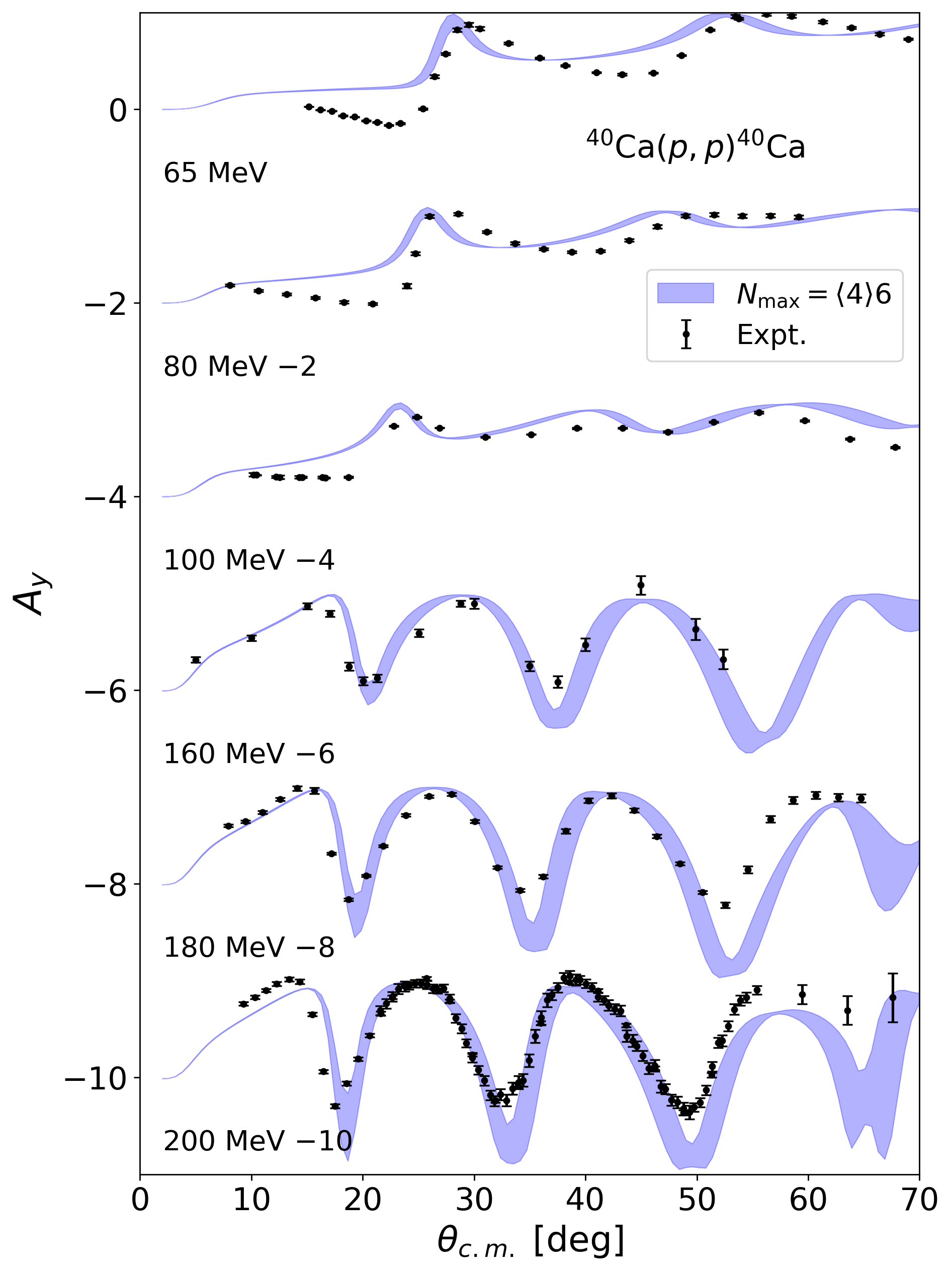}
\caption{
The angular distribution of the analyzing power ($A_y$) for elastic
proton scattering on $^{40}$Ca from 65 to 200~MeV laboratory projectile energy. The additive offset for $A_y$ is given adjacent to the energies listed in the figure.
The calculations use the NNLO$_{\mathrm{opt}}$ interaction in the SA-NCSM with $\left <4\right>6$ model spaces and $\varepsilon_{\mathrm{max}} = 0.03$ (constructed from
$w_{\mathrm{norm}} > 10^{-2}$), and in the leading order in the effective $NA$ spectator interaction. The band indicates variations in the results from $\hbar\Omega=11-13$ MeV. The
data are taken from Ref.~\cite{Sakaguchi:1979fpk} for 65~MeV, Ref.~\cite{Schwandt:1982py} for 80, 160, and
180~MeV, Ref.~\cite{Seifert:1990} for 100~MeV, and 
Refs.~\cite{Stephenson1985,Seifert:1993zz} for 200~MeV.
}
\label{fig8}
\end{center}
\end{figure}

For a careful study of the energy dependence of the elastic scattering observables, we show in Fig.~\ref{fig8} the analyzing powers at the same energies given in Fig.~\ref{fig7}. Here, we observe that the calculations match the minima and maxima of the experimental data well in the entire energy range shown in the figure. However,
for the energies from 100~MeV and below, the experiment shows almost no analyzing power for small angles (momentum transfers),
a feature that is not captured by the calculations, while above 100~MeV, the calculations describe the
experiment very well. Similar to the differential cross section, the agreement of the analyzing power with the experiment is almost perfect at $100$ MeV.

\section{Conclusions and Outlook}
\label{sec:conclusions}

In this work, we concentrate on pursuing the theoretical description of the leading order term of the spectator expansion of the multiple scattering theory that employs
SA-NCSM structure calculations and the first explorations of the SA-NCSM model space selection for scattering observables. This
allows the consideration of heavier nuclei. In this selection procedure, all basis states are kept up to a given $N$, while for higher $N$, the model space is up-selected systematically using \SpR{3} symmetry considerations. This procedure was successfully applied in considering structure phenomena across intermediate- and medium-mass nuclei and is now applied in the context of the construction of {\it ab initio} leading order effective interactions for elastic $NA$ scattering. This effective interaction treats the $NN$ interaction
in the reaction part of the calculation on the same footing as in the structure part. This means that the leading order of the spectator expansion does not only take into account the spin of the projectile nucleon but also that of the struck target nucleon~\cite{Burrows:2020qvu}.
Since this work concentrates on advancing the theoretical description towards heavier nuclei, we only use a single 
chiral potential, namely, NNLO$_{\rm opt}$~\cite{Ekstrom13}.

Because our  work is the first application of the 
selection procedure in the SA-NCSM calculations to scattering,
we first thoroughly tested it by calculating scattering observables for $^6$He. 
We chose this nucleus because scattering calculations using NCSM results with large model spaces have previously been studied~\cite{Burrows:2020qvu}. In addition, $^6$He is a light nucleus, 
but not closed-shell one as is $^4$He.
After establishing the selection procedure, we calculate elastic proton scattering observables for $^{20}$Ne and $^{40}$Ca. 
Though for $^{20}$Ne scattering observables are only available for 65~MeV, it appears that for the chosen 
NNLO$_{\rm opt}$ chiral NN interaction the open-shell $^{20}$Ne is overall slightly better described than the closed shell
$^{40}$Ca at the same energy.
This may be an indication that for deformed nuclei rescattering contributions are less important than for closed-shell
spherical  nuclei.
For $^{40}$Ca we study observables in the energy range from 65 to 200~MeV. Our calculations for differential cross sections and spin observables compare mostly favorably to the experiment. 
We find that the leading order term in the specator lacks in capturing the energy dependence of the differential cross section in forward direction as well as the energy dependence of the first diffraction minimum. The latter was also observed in Ref.~\cite{Vorabbi:2023mml}. The diffraction pattern can be influenced by higher order terms in the multiple scattering series. Therefore, including those would be constructive for theoretical advances.
Summarizing, our study paves the way for applying the SA-NCSM together with a selection procedure that only includes the nonnegligible configurations from the larger model spaces to calculations of the leading order effective $NA$ interaction for nuclei 
with masses around $A\sim 40$-$50$.  Further investigation of medium mass open-shell nuclei 
may give a clearer indication of limitations or successes of the leading order
term of the spectator expansion.


\begin{acknowledgments}
This work was partly performed under the auspices of the U.~S.~Department of Energy under contract Nos.~DE-FG02-93ER40756 and DE-SC0023532, and by the Czech Science Foundation (22-14497S). We thank Daniel Langr for his invaluable contributions to code development. 
The numerical computations benefited from computing resources provided by the Louisiana Optical Network Initiative and HPC resources provided by LSU, together with resources of the National Energy Research Scientific Computing Center, a U.~S.~DOE Office of Science User Facility located at Lawrence Berkeley National Laboratory, operated under contract No.~DE-AC02-05CH11231.
\end{acknowledgments}


\bibliographystyle{apsrev4-1}
\bibliography{lsu_latest,clusterpot,ncsm}


\end{document}